\documentclass{aa}
\usepackage{graphicx}
\usepackage{txfonts}
\usepackage{natbib}
\bibpunct{(}{)}{;}{a}{}{,}
\begin{document}
   \title{The spatial structure of the $\beta$ Pictoris gas disk
   \thanks{Based on observations collected at the European Southern Observatory, Chile.}}


   \author{A. Brandeker\inst{1}
          \and
	  R. Liseau\inst{1}
	  \and
	  G. Olofsson\inst{1}
	  \and
	  M. Fridlund\inst{2}
          }

   \offprints{A. Brandeker}

   \institute{
     Stockholm Observatory, AlbaNova University Centre,
     SE-106 91 Stockholm, Sweden\\
     \email{alexis@astro.su.se, rene@astro.su.se, olofsson@astro.su.se}
     \and
     ESA/ESTEC, PO Box 299, 2200\,AG Noordwijk, The Netherlands\\
     \email{Malcolm.Fridlund@esa.int}
   }

   \date{Received date; accepted date}

   \abstract{ We have used VLT/UVES to spatially resolve the gas disk of
     $\beta$\,Pictoris. 88 extended emission lines are observed, with the
     brightest coming from Fe\,I, Na\,I and Ca\,II. The extent of the gas
     disk is much larger than previously anticipated; we trace Na\,I radially
     from 13\,AU
     out to 323\,AU and Ca\,II to heights of 77\,AU above the disk plane, 
     both to the limits of our observations. The degree of flaring is 
     significantly larger for the gas disk than the dust disk. A strong
     NE/SW brightness
     asymmetry is observed, with the SW emission being abruptly truncated at 
     150--200\,AU. The inner gas disk is tilted about 5\degr\ 
     with respect to the outer disk, similar to the appearance of the disk
     in light scattered from dust. We show that most, perhaps all, of the
     Na\,I column density seen in the 'stable' component of absorption,
     comes from the extended disk. Finally, we discuss the effects of 
     radiation pressure in the extended gas disk and show that the assumption of
     hydrogen, in whatever form, as a braking agent is inconsistent with observations.

   }

   \maketitle
%

\section{Introduction}
The young, near-by main-sequence star $\beta$\,Pictoris has been
the subject of intense studies ever since it was discovered to
harbour circumstellar cold dust \citep{aum85}, distributed along a
linear shape \citep{smi84}, interpreted as a ``debris disk'' 
\citep{bac93}. These studies have been largely motivated by the
possibility of observing an analogue to the solar system in its
early stages, in the hope of finding clues to the mechanisms of
planet formation. Asymmetries found in the disk from light
{\it scattered} by the dust \citep{kal95,hea00}
have indeed been suggested to be the signature of perturbing
planet(s) \citep{mou97,aug01}, but so far no direct detection of
a planet around $\beta$\,Pic has been made.
Asymmetries have also been detected in {\it thermal emission}
from the dust (\citealt{lis03b}; \citealt{wei03} with references
therein). Recent papers reviewing the $\beta$\,Pic disk are those
by \citet{art00}, \citet{lag00} and \citet{zuc01}.

Circumstellar {\it gas}, seen in absorption against
the star, was also found early on \citep{hob85}, thanks to the
favourable edge-on orientation of the disk. Finding and characterising
the gas content is important for understanding its relation to the
dust and the general evolution of the disk \citep{art00}. Gas
is also useful as a probe of physical conditions in the disk,
where density, composition, temperature and bulk velocities 
under favourable conditions can be directly estimated.

The gas found at relative rest to $\beta$\,Pic, consisting
of metals, raised the problem why it is not blown away from the
system by the high radiation pressure.
\citet{lag98} made some detailed calculations and found that the
gas drag from a dense enough H\,I ring
($n_{\mathrm{H\,I}} \ge 10^5$\,cm$^{-3}$) close to the star
($\sim$0.5\,AU) could brake migrating particles sufficiently,
provided they started out inside the ring.
The picture was complicated by the announcement of H$_2$
detected in emission by the Infrared Space Observatory \citep{thi01},
implying large quantities ($\sim$50\,M$_{\oplus}$) of molecular
hydrogen, and the sub-sequent report of sensitive upper limits 
($N(\mathrm{H_2}) \la 10^{18}$\,cm$^{-2}$) of H$_2$ from
FUV absorption lines using $\beta$\,Pic as a
background source \citep{lec01}. In addition, 
\citet[ hereafter Paper\,I]{olo01} found spatially resolved widespread
gas emission from Na\,I in the disk, stretching out to at least 140\,AU.

Here we present observations improved by a factor of two in both spatial
and spectral resolution, as well as a greatly increased spectral coverage,
compared to Paper\,I.
We put emphasis on the observed spatial structure of the gas disk,
derive an empirical density profile of Na\,I atoms and use a
photoionisation code to construct disk models consistent with our
observations. We discuss implications of the radiation pressure under
various conditions derived from these models. Results from a detailed
study of the observed chemical abundances will be discussed in a
forthcoming paper.

\section{Observations}
\begin{figure}
 \resizebox{\hsize}{!}{\includegraphics{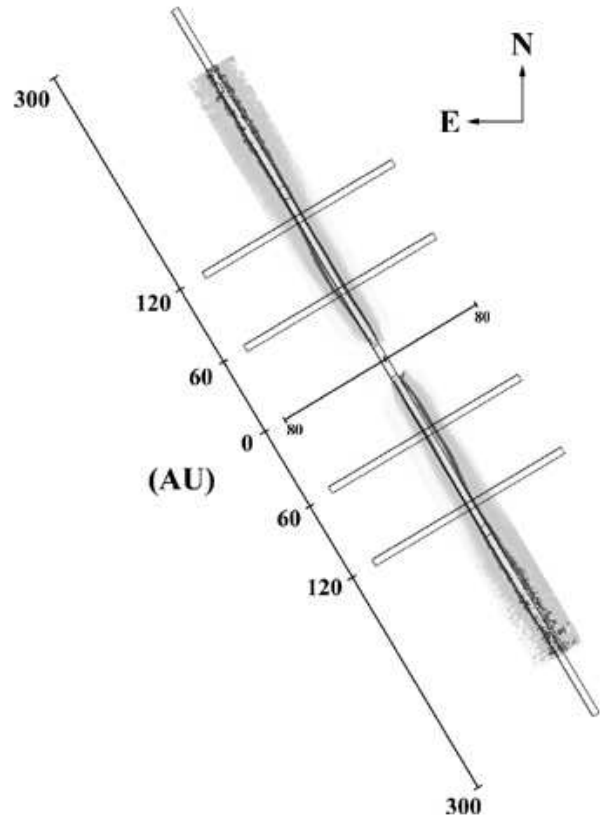}}
 \caption{Orientations and positions of the 8 different slit settings used. The background
  image shows observations of the dust as obtained by HST/STIS \citep{hea00}.}
 \label{slitpos}
\end{figure}

The star $\beta$~Pictoris was observed with the echelle spectrograph
UVES on the 8.2\,m Kueyen telescope at the Very Large Telescope of ESO,
Paranal, Chile. The observations were made in service mode to take advantage
of the excellent seeing conditions occasionally provided by the site.
Service mode means that the observers prepare observation blocks that
are executed by the local telescope operators provided certain
conditions (like air mass and seeing) are met. A total of 12 
observation blocks during 2001--2002 (see Table~\ref{obslog}) lasting about an
hour each were successfully executed. The air mass and seeing at zenith
were always below 1.5 and 0\farcs6, respectively. The UVES spectrograph
was used in standard dichroic modes, meaning that a dichroic mirror was used
to split the light into two wavelength ranges, each using its own 
echelle grating and CCD detectors. Two types of gratings were used for
each arm, together covering the wavelengths 3\,300\,\AA\ to
10\,400\,\AA\ with a spectral resolution ranging from $R$$\sim$80\,000 in the blue
to $R$$\sim$110\,000 in the red. The blue arm was equipped with a single CCD of
type EEV~44-82 and size 4\,096$\times$4\,096 pixels, while the red arm had
a mosaic of two CCDs of sizes 2\,048$\times$4\,096 each, of the types
EEV~44-82 and MIT-LL CCID-20. The spectroscopic slits were 0\farcs4 and 
0\farcs3 wide in the blue and red arm, respectively. To adequately separate
the echelle orders, slits were restricted to
lengths between 8\arcsec\ and 12\arcsec, depending on setting
(see Table~\ref{obslog}). Slits were positioned at 4 overlapping locations along 
the $\beta$\,Pic dust disk, and at 4 locations orthogonal to the
disk, reaching a distance of 17\arcsec\ from the star in the disk plane, and
a height of 6\arcsec\ above the plane (see Fig.~\ref{slitpos}). From the data,
we determined the positioning of the slits, with respect to $\beta$\,Pic, to
be better than $\sim$0\farcs2. Calibration data (flat-fields, bias frames,
Th-Ar lamp wavelength calibration spectra, flux standard stars, etc.) were
provided by the UVES calibration plan.

\begin{table*}
 \caption[]{VLT/UVES observation log}
 \label{obslog}
 \begin{tabular}{cccccccc}
   \hline
   UT Date & $\beta$\,Pic$^{\mathrm{a}}$ & Grating$^{\mathrm{b}}$ & Width$^{\mathrm{c}}$
   & Length$^{\mathrm{c}}$ & Offset$^{\mathrm{d}}$ & Orientation$^{\mathrm{d}}$ & Exposures \\
   \hline
   2001-12-02\,T\,03:38 & ON & 437 & 0\farcs4 & 10\arcsec & -4\farcs4 & Parallel & 69 $\times$ 15\,s \\
   2001-12-02\,T\,03:38 & ON & 860 & 0\farcs3 & 12\arcsec & -4\farcs4 & Parallel & 47 $\times$ 25\,s \\
   2001-12-02\,T\,05:31 & ON & 390 & 0\farcs4 &  8\arcsec & -3\farcs4 & Parallel & 62 $\times$ 20\,s \\
   2001-12-02\,T\,05:31 & ON & 580 & 0\farcs3 & 12\arcsec & -3\farcs4 & Parallel & 48 $\times$ 25\,s \\
   2002-01-03\,T\,06:21 & ON & 390 & 0\farcs4 &  8\arcsec & -3\farcs4 & Parallel & 62 $\times$ 20\,s \\
   2002-01-03\,T\,06:21 & ON & 580 & 0\farcs3 & 12\arcsec & -3\farcs4 & Parallel & 48 $\times$ 25\,s \\
   2002-01-06\,T\,05:51 & OFF & 390 & 0\farcs4 &  8\arcsec & -3\farcs0 & Orthogonal & 3 $\times$ 1205\,s \\
   2002-01-06\,T\,05:51 & OFF & 580 & 0\farcs3 & 12\arcsec & -3\farcs0 & Orthogonal & 3 $\times$ 1195\,s \\
   2002-01-16\,T\,02:47 & OFF & 390 & 0\farcs4 &  8\arcsec & -11\arcsec & Parallel & 3 $\times$ 1205\,s \\
   2002-01-16\,T\,02:47 & OFF & 580 & 0\farcs3 & 12\arcsec & -11\arcsec & Parallel & 3 $\times$ 1195\,s \\
   2002-01-16\,T\,03:51 & OFF & 390 & 0\farcs4 &  8\arcsec & -6\farcs0 & Orthogonal & 3 $\times$ 1205\,s \\
   2002-01-16\,T\,03:51 & OFF & 580 & 0\farcs3 & 12\arcsec & -6\farcs0 & Orthogonal & 3 $\times$ 1195\,s \\
   2002-01-19\,T\,03:11 & ON & 390 & 0\farcs4 &  8\arcsec & 3\farcs4 & Parallel & 62 $\times$ 20\,s \\
   2002-01-19\,T\,03:11 & ON & 580 & 0\farcs3 & 12\arcsec & 3\farcs4 & Parallel & 48 $\times$ 25\,s \\
   2002-01-19\,T\,04:25 & ON & 390 & 0\farcs4 &  8\arcsec & 3\farcs4 & Parallel & 62 $\times$ 20\,s \\
   2002-01-19\,T\,04:25 & ON & 580 & 0\farcs3 & 12\arcsec & 3\farcs4 & Parallel & 48 $\times$ 25\,s \\
   2002-02-09\,T\,03:55 & OFF & 390 & 0\farcs4 &  8\arcsec & 3\farcs0 & Orthogonal & 3 $\times$ 1205\,s \\
   2002-02-09\,T\,03:55 & OFF & 580 & 0\farcs3 & 12\arcsec & 3\farcs0 & Orthogonal & 3 $\times$ 1195\,s \\
   2002-02-10\,T\,03:19 & OFF & 390 & 0\farcs4 &  8\arcsec & 11\arcsec & Parallel & 3 $\times$ 1205\,s \\
   2002-02-10\,T\,03:19 & OFF & 580 & 0\farcs3 & 12\arcsec & 11\arcsec & Parallel & 3 $\times$ 1195\,s \\
   2002-02-11\,T\,04:04 & OFF & 390 & 0\farcs4 &  8\arcsec & 6\farcs0 & Orthogonal & 3 $\times$ 1205\,s \\
   2002-02-11\,T\,04:04 & OFF & 580 & 0\farcs3 & 12\arcsec & 6\farcs0 & Orthogonal & 3 $\times$ 1195\,s \\
   2002-09-24\,T\,08:09 & OFF & 437 & 0\farcs4 & 10\arcsec & -3\farcs0 & Orthogonal & 3 $\times$ 1205\,s \\
   2002-09-24\,T\,08:09 & OFF & 860 & 0\farcs3 & 12\arcsec & -3\farcs0 & Orthogonal & 3 $\times$ 1195\,s \\
   \hline
 \end{tabular}
 \begin{list}{}{}
 \item[$^{\mathrm{a}}$] ON means that the star was inside the spectroscopic slit, OFF that it was outside.
 \item[$^{\mathrm{b}}$] The gratings correspond to the following wavelength ranges:
   326--445\,nm for grating 390,
   373--499\,nm for grating 437,
   476--684\,nm for grating 580, and
   660--1060\,nm for grating 860.
 \item[$^{\mathrm{c}}$] Width and Length refer to the dimensions of the slit.
 \item[$^{\mathrm{d}}$] Offset is the centre of the slit relative to $\beta$\,Pic in the plane
 of the disk, where positive offsets are to the north-east. The orientation of the
 slit is relative to the disk plane.
 \end{list}
\end{table*}

\section{Data reduction}
\subsection{Pipeline reduction}
The data were reduced using a modified version of the UVES pipeline 1.2,
that runs in the ESO MIDAS environment. The pipeline
automatically generates the appropriate calibration sets from the
calibration data obtained as a part of the UVES calibration plan,
and apply them to the science frames. The raw frames are bias
subtracted, flat-fielded, background subtracted, order extracted,
and wavelength calibrated. An error was corrected in the pipeline
version 1.2, that associated a wavelength calibration to the wrong
order when 2D extraction was selected without merging the orders.

The orders of the echelle grating are slightly inclined on the CCD
with respect to pixel columns. Since the observations with the
star on slit were performed under excellent seeing (the seeing
measured in the spectra was $\sim$0\farcs7), this
resulted in a very high flux gradient between the pixels on and
off the star. By default, the orders are resampled linearly,
but this caused the spectrum to show a periodic 'spiky' pattern
in the flux. We found that linear interpolation does not work
properly due to the strong
non-linear shape of the point spread function (PSF) at the pixel
resolution. By modifying the order extracting routine to use a third
order Catmull-Rom spline interpolation instead of linear, we found
the spikes to be greatly reduced (though not completely eliminated).

The reduction procedure was complicated by the sheer load of
data. With about 600 spectra, each 4\,096$\times$4\,096 pixels,
plus calibration frames etc., the raw data were close to 50\,GB.

The end product of the pipeline reduction consisted of
individual orders separated into individual files with 
the spatial information along the slit preserved, and a
wavelength calibration guaranteed to be better than
0.5\,km\,s$^{-1}$.

\subsection{PSF subtraction}
To trace the circumstellar gas emission as close as possible to
the star, we had to subtract the scattered light from the star,
in the cases where the star was in the slit. To do this, we
estimated the stellar spectrum $S$ by centering a small aperture
on the star. To allow for slow gradients, we took advantage of the
fact that the gas emission lines are narrow (a few pixels) and
produced a median filtered version $M$ of the stellar spectrum
$S$, using a window of 30--60 pixels. We also obtained the median
filtered spectrum $m_i$ at each spatial position $i$, using the
same median filter window. The stellar PSF subtracted spectrum was
then estimated as $s_i = r_i - m_i S/M$,
where $r_i$ is the spectrum at position $i$. For the observations
where the star was off the slit, we just subtracted the median filtered
background scattered light (scattered both locally by the dust
disk and in the atmosphere / telescope), $s_i = r_i - m_i$.

\subsection{Flux calibration}
To flux-calibrate the data, we made use of the master response curves 
provided by the UVES team. The master response curves are generated
from long-time monitoring of the sensitivity trends of the instrument,
and are provided for various standard settings and periods in time.
The claimed absolute flux calibration is 10\,\%, but comparing with
standard flux stars observed (with an open slit) the same nights as
$\beta$\,Pic we found the derived fluxes to deviate by as much as 40\,\% from
tabulated values for these stars. We thus used the master response
curves to correct only for the over-all sensitivity dependence on wavelength,
and used the standard star spectra obtained in connection with the
$\beta$\,Pic observations to correct the absolute flux calibration. By
looking at the variations of the derived absolute fluxes of disk emission
lines observed several times, we estimate that the error on the absolute
flux is about 5--10\,\%. Since we use a narrow slit (0\farcs3--0\farcs4),
the slit losses from a point source like $\beta$\,Pic are substantial and
very seeing dependent. The error of the absolute flux from the star is
therefore expected to be substantially higher. Our main concern, however,
is the flux of the disk emission, which, due to its spatial extension, is
much more stable with regard to seeing variations.

\subsection{Heliocentric wavelength correction}
The wavelength calibration obtained from the UVES pipeline does not
correct for the velocity of the instrument relative to the centre of
the Sun. To transform the observed spectrum to a heliocentric frame,
we made use of the software program RV written by P.\ T.\ Wallace and
C.\ A.\ Clayton\footnote{RV is available from http://star-www.rl.ac.uk}.
The obtained accuracy in the transformation is better
than 0.01\,km\,s$^{-1}$, fully adequate for our purposes.

\subsection{Line flux measurements}
\label{linemeasure}
From the reduced, PSF subtracted and wavelength calibrated spectra,
the disk emission lines were measured in several steps. First the
lines were identified by correlating their measured wavelengths with 
the atomic line database provided by NIST\footnote{http://physics.nist.gov}. 
This procedure
was greatly simplified by the high accuracy of the wavelength calibration,
better than 0.5\,km\,s$^{-1}$, corresponding to 0.01\,\AA\ at 6\,000\,\AA. A
median systematic velocity was then calculated for the brightest and most
accurately measured emission lines, assuming all shared the same systematic
radial velocity. For the brightest lines, the measured flux as a function
of apertures centred on the systematic velocity was evaluated. A large
aperture samples more signal, but also more noise. We were therefore
interested in finding the best balance, in order to achieve the highest
signal to noise ($S/N$). The apertures used to measure emission lines were
consequently chosen as a function of line strength and background noise,
with smaller apertures for fainter lines. We assumed that all lines from
a particular ion share the same spatio-spectral profile, and scaled the flux 
measured in small apertures of fainter lines with the ratio between equally
sized and maximum sized apertures placed on the sum of several bright lines.
In this way we estimated the signal from a faint line without integrating up
too much noise. The method is analogous to methods used in aperture
photometry to measure star fluxes in, e.g., a CCD image.

For the slits placed orthogonally to the disk, we summed up all flux in the
spatial direction, i.e. along the height of the disk (Table~\ref{emitab}).
The quantity thus derived has the unit of flux per arcsecond. 

\section{Results}
\begin{figure}
 \resizebox{\hsize}{!}{\includegraphics{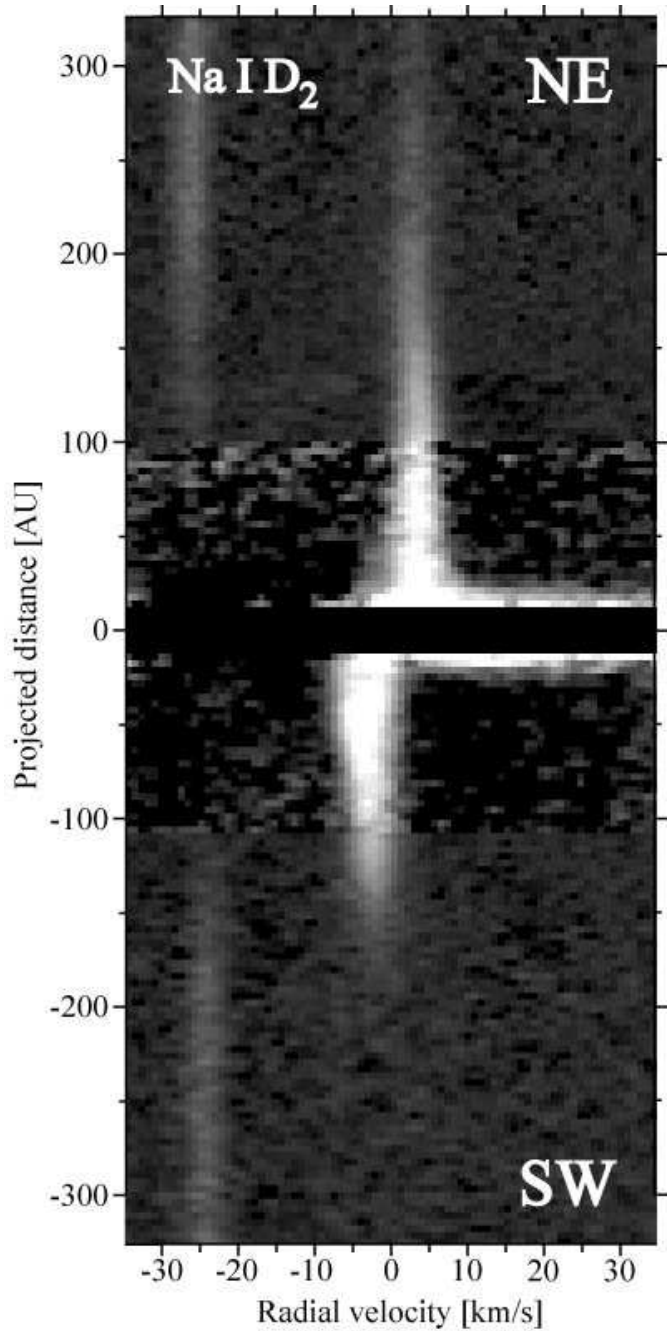}}
 \caption{Na\,D$_2$ ($\lambda_{\mathrm{air}}$ = 5889.951\,\AA) emission from the
   $\beta$\,Pic disk, as seen through
   four slits parallel to the disk (see Fig.~\ref{slitpos}). The vertical
   axis is along the spatial direction, with positive offsets
   north-east of the star, while the spectral dispersion is along the
   horizontal axis, centred on the $\beta$\,Pic rest frame of the
   Na\,D$_2$ line. The inner 11\,AU, showing mostly residual noise from the 
   PSF subtraction, have been masked out. The velocity shift of the NE and
   SW side is due to Keplerian rotation of the disk, with the SW rotating
   towards us (Paper\,I). The grey scale
   has been scaled non-linearly with the intensity, in order to bring out
   the bright disk structure close to the star as well as the faint features
   at greater distances. The emission in the NE can be traced out to the
   limits of our observations at 323\,AU, while the emission in the SW ends
   abruptly at 150--200\,AU (see also Fig.~\ref{radial}).
   Close to $-$25\,km\,s$^{-1}$, telluric Na\,D$_2$ emission is seen covering
   the slit. The relative velocity of the sky emission to $\beta$\,Pic varies
   slightly between the mosaiced observations due to the orbital motion of
   the Earth between the epochs.}
 \label{spatspec}
\end{figure}

\begin{figure*}
 \resizebox{\hsize}{!}{\includegraphics{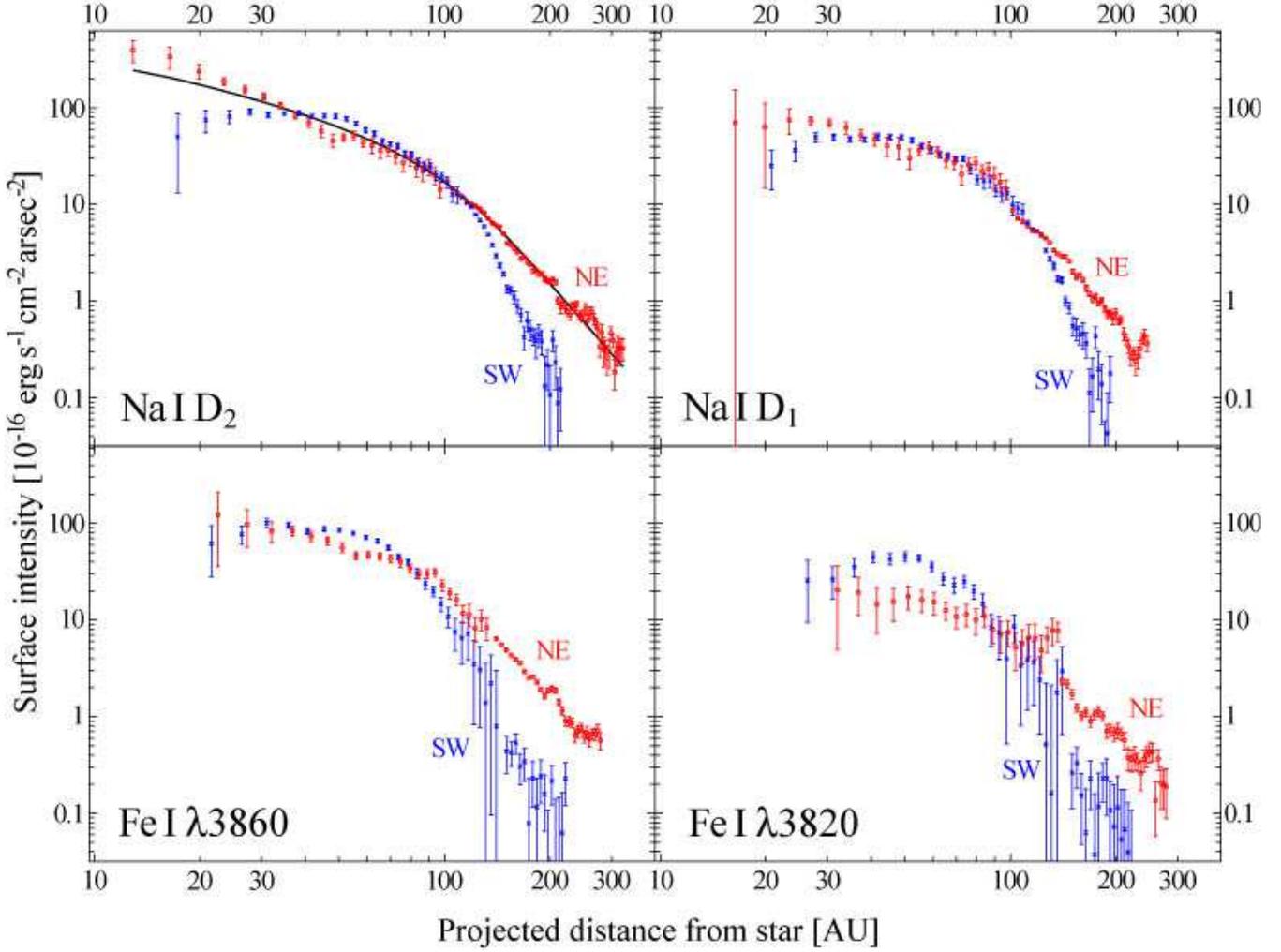}}
 \caption{Determined surface intensity as a function of radial distance in NE and SW for
Na\,D$_{1,2}$ and Fe\,I $\lambda$3859.911,\,$\lambda$3820.425. The fit of
Sect.~\ref{denssect} is shown for Na\,I\,D$_2$ as a an unbroken line.
Note the sharp decrease in flux in the SW at 150\,AU--200\,AU.}
 \label{radial}
\end{figure*}

We detected 88 spatially extended emission lines from
the $\beta$\,Pic gas disk, identified as emission from
Fe\,I, Na\,I, Ca\,II, Ni\,I, Ni\,II, Ti\,I, Ti\,II, Cr\,I and
Cr\,II. Table~\ref{emitab} shows a selection of the brightest lines.
Following the brightest emission lines 
($S/N$$\sim$50) from Na\,I and Fe\,I radially (Figs.~\ref{spatspec}~\&~\ref{radial}), we
observe a strong asymmetry between the north-east (NE) and the south-west
(SW) parts of the disk, similar to the brightness asymmetry in the dust
emission \citep{kal95},
but asymmetric to a much higher degree. The NE gas emission extends
smoothly to the limits of our observations (17\arcsec, corresponding to 
330\,AU at the distance of $\beta$\,Pic), whereas the SW emission is
abruptly truncated at 150--200\,AU. In the inner
regions, the SW emission dominates over the NE part, in agreement with
Fig.~2 of Paper\,I.

\begin{table}
 \caption[]{Selected gas emission lines, 3\arcsec\ SW in disk}
 \label{emitab}
 \begin{tabular}{lcrccc}
   \hline
   Line & $\lambda_{\mathrm{air}}^{\mathrm{a}}$ &
   Flux$^{\mathrm{b}}$ &
   $\sigma_{\mathrm{flux}}^{\mathrm{b}}$ &
   $E_{\mathrm{low}}^{\mathrm{c}}$ &
   $E_{\mathrm{high}}^{\mathrm{c}}$ \\
   \hline
   Fe\,I        & 3820.425 & 61.3  & 2.5 & 0.859 & 4.103 \\
   Fe\,I        & 3859.911 & 111.3 & 3.4 & 0.000 & 3.211 \\
   Na\,I~D$_2$  & 5889.951 & 79.2  & 4.0 & 0.000 & 2.104 \\
   Na\,I~D$_1$  & 5895.924 & 42.4  & 2.5 & 0.000 & 2.102 \\
   Ca\,II~K     & 3933.663 & 12.0  & 1.0 & 0.000 & 3.151 \\
   Ca\,II~H     & 3968.468 & 16.8  & 1.2 & 0.000 & 3.123 \\
   \hline
 \end{tabular}
 \begin{list}{}{}
 \item[$^{\mathrm{a}}$] Wavelength of transition in air, in units of \AA.
 \item[$^{\mathrm{b}}$] Flux in units of 
   10$^{-16}$\,erg\,s$^{-1}$cm$^{-2}$arcsec$^{-1}$; see Sect.~\ref{linemeasure}.
 \item[$^{\mathrm{c}}$] Lower and upper energy levels of transition, in units of eV.
 \end{list}
\end{table}

\begin{figure}
 \resizebox{\hsize}{!}{\includegraphics{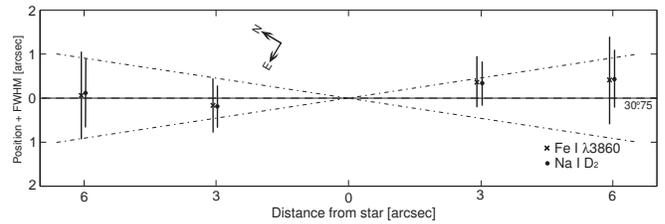}}
 \caption{Measured positions of the Na\,I and Fe\,I gas emission
relative to the plane at position angle 30\fdg75. The bars correspond
to derived scale heights, obtained by deconvolving the observed height
profiles with a Gaussian of 0\farcs7 FWHM simulating the seeing. The
inclined dash-dotted lines show the empirical opening angle of the gas
disk, corresponding to $H/r=0.28$ (see Sect.~\ref{denssect}).
The error in positioning of the slits is on the order of 0\farcs1.}
 \label{fwhmpos}
\end{figure}

Determining the centre of emission from the orthogonal profiles,
it becomes apparent that the inner part of the disk is slightly
tilted with respect to the outer parts, in particular on the
NE side (Fig.~\ref{fwhmpos}). We estimate this
tilt to be 5\degr $\pm$2\degr\ (1$\sigma$), similar to the
4\degr --5\degr\ tilt observed by HST/STIS in the inner dust
disk \citep{hea00}.

\begin{figure*}
\resizebox{!}{23 cm}{\includegraphics{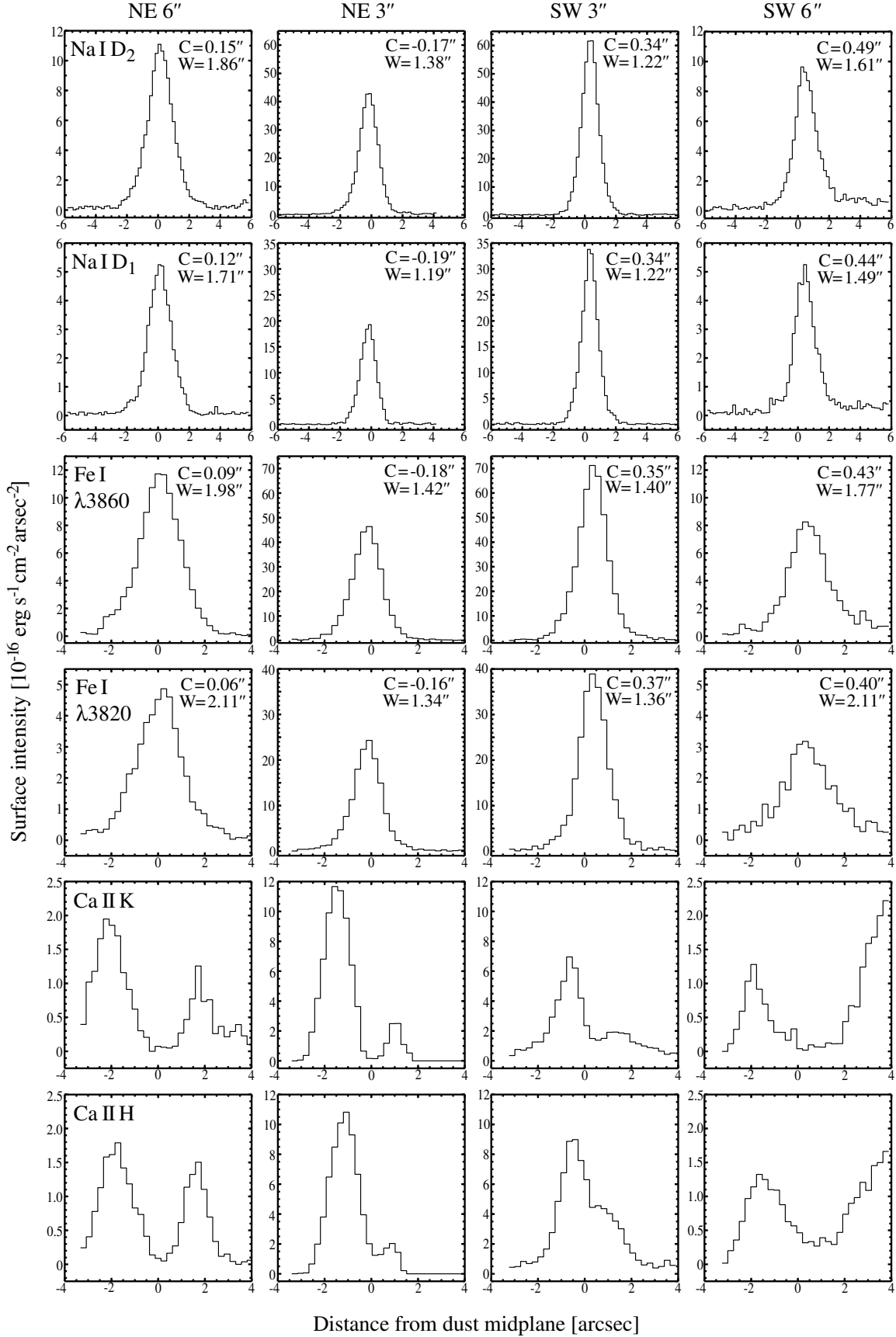}}
\caption{The flux calibrated vertical profile of
the disk gas emission at 3\arcsec\ and 6\arcsec\ NE and SW of the
star for a few selected lines of Na\,I, Fe\,I and Ca\,II.
Zero spatial offset refers to the dust disk midplane defined
by the position angle 30\fdg75 relative to the star, with
the positive spatial direction to the north-west (NW). The
centre and FWHM for Gaussian fits to the profiles are printed
in the panels of Gaussian shaped profiles. The SW
side dominates the emission at 3\arcsec, while the NE side
slightly dominates at 6\arcsec.}
\label{orthogonal}
\end{figure*}

The scale height (FWHM) of the gas disk, estimated on observations deconvolved
with a Gaussian of 0\farcs7 to simulate the seeing, is $\sim$20\,AU at 3\arcsec\ (58\,AU),
similar to the dust disk scale height \citep{hea00}. At 6\arcsec\ (116\,AU), however,
the gas disk is significantly thicker, $\sim$30\,AU compared to $\sim$15\,AU
for the dust (see Fig.~\ref{orthogonal}).

\begin{figure*}
 \resizebox{\hsize}{!}{\includegraphics{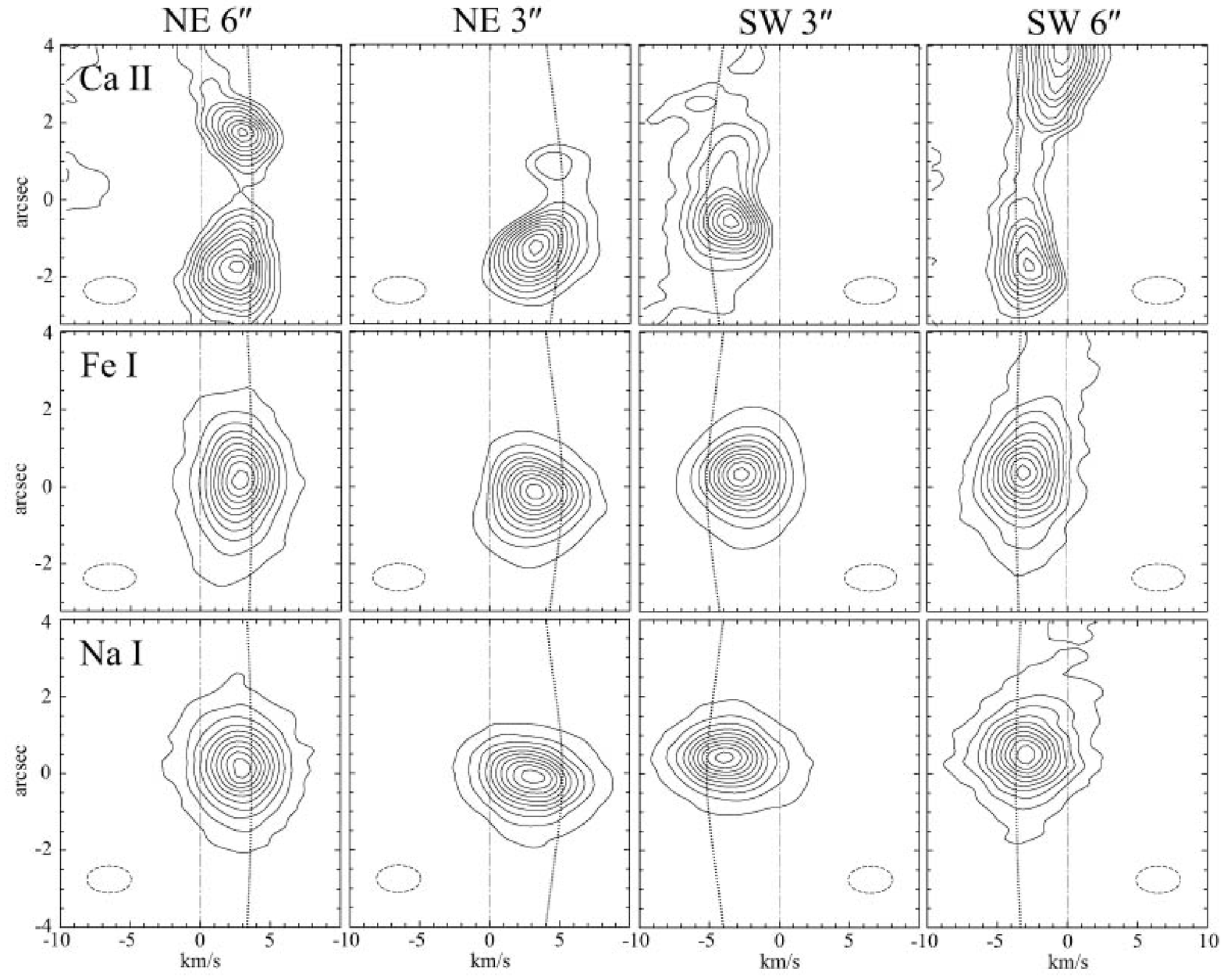}}
 \caption{The velocity distribution of the gas
emission as observed by slits orthogonal to the disk at 3\arcsec\ 
and 6\arcsec\ NE and SW of the star for Ca\,II, Fe\,I and Na\,I.
To improve the signal to noise, the H \& K lines of Ca\,II,
10 of the brightest Fe\,I lines, and D$_1$ and D$_2$ of  Na\,I
were added for each species. The profiles were then normalised so
that each level increases one tenth of the peak. The noise in the
Fe\,I and Na\,I plots is below the lowest level curve, while the
lowest level curve for Ca\,II still shows some noise.
Velocities are referred to the
system velocity 20.0\,km\,s$^{-1}$ of $\beta$\,Pic, with the positive
direction being away from us. Positive offsets in the spatial direction
are directed to the north-west (NW). The dash-dotted vertical lines
correspond to the systematic velocity of $\beta$\,Pic, the dotted lines
to the Kepler velocity at the projected distance from $\beta$\,Pic
(assuming a stellar mass of $M = 1.75\,M_{\odot}$),
while the dashed ellipse in each graph shows the spectral and
spatial resolution attained. Note the complexity of
the Ca\,II profile, and how the emission at 6\arcsec\ SW
(right-most panels) can be traced to the limit of the
observations. The detected emission profiles from Ni\,I, Ni\,II,
Ti\,I, Ti\,II, Cr\,I and Cr\,II cannot be distinguished from 
Na\,I or Fe\,I, due to the lower S/N. }
 \label{contours}
\end{figure*}

A very complex height profile is shown by the Ca\,II H \& K lines
(Figs.~\ref{orthogonal}~\&~\ref{contours}). Especially at 6\arcsec\ 
distance from the star, the emission from the disk midplane is much
fainter than the emission away from the midplane. In particular,
the detected Ca\,II emission, at 6\arcsec\ SW of the star, keeps
increasing to the limits of the spectroscopic slit 4\arcsec\ above
the disk midplane, meaning that there is a significant number of
Ca\,II ions at 77\,AU height above the midplane at 116\,AU distance
from the star. From Fig.~\ref{contours} it is evident that the
radial velocities of these Ca\,II ions are small -- on the order
of a few km\,s$^{-1}$. Implications of this fact are discussed
in Sect.~\ref{radpress}.

In order to determine the heliocentric velocity of $\beta$\,Pic
accurately, we averaged the heliocentric gas velocities as
observed on each side at the distances 3\arcsec\  and 6\arcsec\ 
from the star. In this way, we can assess the system velocity
independently from the lines observed in absorption, which are
sensitive to possible radial velocities of the absorbers
(caused by, e.g., radiation pressure). We determined the
heliocentric radial velocity of $\beta$\,Pic to be
20.0$\pm$0.5~km\,s$^{-1}$, with the error mainly due to the
uncertainty in wavelength calibration.

In the spectrum of the star, we measured the absorption profiles
corresponding to observed ground state emission lines. Most lines
were found to be at, or close to, the systematic velocity of
$\beta$\,Pic (Table~\ref{radveltab}).

\begin{table}
 \caption[]{Radial velocities relative $\beta$\,Pic of ions measured in absorption}
 \label{radveltab}
 \begin{tabular}{lcc}
   \hline
   Ion & $\Delta v^{\mathrm{a}}$ &
   $\sigma_{\Delta v}^{\mathrm{b}}$ \\
       & km\,s$^{-1}$ & km\,s$^{-1}$ \\
   \hline
   Fe\,I  & $-$0.0 & 0.3 \\
   Na\,I  & $-$1.2 & 0.3 \\
   Ca\,II$^{\mathrm{c}}$ & $-$0.3 & 2.1\\
   Ti\,II & 0.2 & 0.8 \\
   Ni\,I  & 0.4 & 0.4 \\
   Ni\,II & 2.8 & 3 \\
   Cr\,II & 2.2 & 3 \\
   \hline
 \end{tabular}
 \begin{list}{}{}
 \item[$^{\mathrm{a}}$] The positive direction is radially away from us, towards $\beta$\,Pic.
 \item[$^{\mathrm{b}}$] Errors are relative only. An additional systematic error of
   0.5\,km\,s$^{-1}$ comes from the uncertainty in the system velocity of $\beta$\,Pic.
 \item[$^{\mathrm{c}}$] The radial velocity of Ca\,II is measured on the infrared lines
 $\lambda$7291.47 and $\lambda$7323.89, and not the optically thick Ca\,II H \& K lines.
\end{list}
\end{table}

\section{Discussion}

We proceed by inverting the observations of the projected disk
to a radial density law for Na\,I, and then extend the result
by calculating the ionisation structure to find an estimate on
the total Na density. By making assumptions of the abundances
we construct two models with distinct H densities,
test how well the models compare with observations of H\,I and
H$_2$, and examine how the kinematics of ions in these models are
affected by radiation pressure.

\subsection{Sodium gas density profile}
\label{denssect}
From the spatial emission profiles we estimate the density distribution
of the observed gas by making the following assumptions:
\begin{enumerate}
\item The disk is axisymmetric and the density of the emitting medium
  is well described by
\begin{equation}
\label{denseq}
n(r,h) = n_0 \left[
\left( \frac{r}{r_0} \right)^{2a} +
\left( \frac{r}{r_0} \right)^{2b}   \right] ^{-\frac{1}{2}}
 \exp \left[ - \left( \frac{h}{\alpha r} \right) ^2 \right], 
\end{equation}
where $r$ and $h$ are the cylindrical coordinates describing the
midplane distance and height over the midplane, respectively, $n_0$
is a normalising density, $r_0$ is the broken power-law break
distance, $a$ and $b$ are power-law exponents of the inner and
outer regions respectively, and $\alpha$ is related to the scale
height $H$ of the disk at midplane
distance $r$ as $H/r = 2\sqrt{\ln 2}\,\alpha$.
\item The inclination of the disk is close to zero, i.e. we
  see it edge-on.
\item The emission is optically thin and isotropic.
\item The emitting gas traces the full gas population of that ion.
\end{enumerate}
The observed disk is, of course, not axisymmetric but quite
asymmetric in appearance
(Figs.~\ref{spatspec},~\ref{radial}~\&~\ref{contours}). The
difference between the NE and SW parts of the inner disk 
is not that dramatic, though, less than a factor of two
in the region where most gas is located (30--120\,AU). Thus
the assumption of axi\-symmetry is probably not too bad.

We chose the Na\,I\,D$_{1,2}$ lines for the inversion because
(A) the $S/N$ is high ($\sim$50), (B) the line ratio
between D$_2$ and D$_1$ is close to 2, implying that the emission
is optically thin, and (C) we expect nearly all Na\,I atoms 
to be in their ground state (see Sect.~\ref{cloudy}), meaning
the resonance D$_{1,2}$ lines trace all of the Na\,I population.
The parameters of Eq.\,\ref{denseq} were fitted by constructing
a numerical model, simulating observed spatial profiles as
a function of the parameters and $\chi ^2$-minimising the 
difference between the model and the observations. In more
detail, the model was initiated on a three dimensional grid
of height $\times$ width $\times$ depth dimensions
40\,AU $\times$ 800\,AU $\times$ 2000\,AU at 1\,AU resolution.
The central star was assumed to emit
4.5$\times$10$^{30}$\,erg\,s$^{-1}$\AA$^{-1}$ at wavelengths
close to the Na\,I\,D$_2$ line, and
$A_{ji}=6.22\times10^7$\,s$^{-1}$ was
used as the D$_2$ transition Einstein coefficient for
spontaneous emission, in calculating the scattering
cross-section. For each cell, a specific line luminosity
was calculated, summed up along the line of sight and
geometrically diluted to the distance of the observer at
19.3\,pc. This 'ideal' image of the disk was then convolved
with a Gaussian of FWHM = 0\farcs7, simulating the atmospheric
seeing, and sampled along the various spectroscopic
slit settings of Fig.~\ref{slitpos} to fit the observations.
In the fit we used the midplane
radial emission distribution from 13\,AU out to 323\,AU of
the NE, and the orthogonal profiles at 3\arcsec\ and 6\arcsec\
NE of the star. The derived parameters of Eq.\,\ref{denseq} for
Na\,I are:
\begin{eqnarray*}
n_0 & = & (1.02\pm0.04)\times 10^{-5}\,\mathrm{cm}^{-3} \\
r_0 & = & 117\pm3~\mathrm{AU} \\
a & = & 0.47\pm0.06 \\
b & = & 3.16\pm0.04 \\
H/r & = & 0.28\pm0.05,
\end{eqnarray*}
where the quoted errors are 1$\sigma$\ formal fitting errors
obtained by making an additional number of fits
to artificial data generated by adding random noise, at the
estimated noise level, to the observed profile. We
have over plotted the Na\,I fit into Fig.~\ref{radial}.

We can use Eq.\,\ref{denseq} with the above parameters to
estimate the midplane column density of Na\,I, to compare
with column densities estimated from Na\,I absorption in
the stellar spectrum seen through the disk. If we 
integrate only over the radii where we are sensitive to
Na\,I emission, that is, from 13\,AU out to 323\,AU, we
obtain a column density of $N(\mathrm{Na\,I}) = 
3.0\times10^{10}$\,cm$^{-2}$.
Extrapolating Eq.\,\ref{denseq} to all radii, that is,
from zero radius out to infinity, we increase the
Na\,I column density by 40\,\% to $N(\mathrm{Na\,I}) =
4.1\times10^{10}$\,cm$^{-2}$. This can be compared to
our observed column density $N(\mathrm{Na\,I}) =
(3.4\pm0.4)\times10^{10}$\,cm$^{-2}$. We conclude that
most, perhaps all, of the Na\,I gas seen in absorption
is situated in the extended gas disk seen in emission.
We have no reason to doubt that originators of other
'stable' gas absorption components also belong to this
extended disk.


\subsection{Radiation pressure}
\label{radpress}
Ions in the $\beta$\,Pic gas disk are subject to appreciable
radiation pressure. To evaluate the significance, it is common
to define the ratio between the forces of gravitation and radiation,
$\beta \equiv F_{\mathrm{rad}} / F_{\mathrm{grav}}$.
Since both the radiation and gravitational fields are inversely
proportional to the square of the distance to the source, $\beta$
is constant throughout the $\beta$\,Pic disk. For the observed
atoms Na\,I and Fe\,I, $\beta \gg 1$, meaning that if left alone,
gravity would be largely irrelevant and these atoms would rapidly
accelerate out of the system at high velocities. This is not observed
(Table~\ref{radveltab} and Fig.~\ref{contours}). To investigate 
what the effects of a braking medium are, we solve the
equation of motion for a gas particle,
\begin{equation}
\label{motioneq}
m\frac{\mathrm{d}v}{\mathrm{d}t} = -F_{\mathrm{grav}} +
   F_{\mathrm{rad}} - F_{\mathrm{fric}},
\end{equation}
where $F_{\mathrm{fric}}$ is the frictional force. Assuming
$F_{\mathrm{fric}} = C v$, where $C$ is the friction coefficient
(proportional to the density of the braking medium), and that
the distance an atom travels before reaching the terminal velocity
is much shorter than the size of the disk, Eq.\,\ref{motioneq}
may be solved to yield the terminal velocity (as $t \rightarrow \infty$)
\begin{equation}
\label{terminaleq}
v_{\infty} = \frac{\beta - 1}{C} F_{\mathrm{grav}}.
\end{equation}
Eq.\,\ref{motioneq} ignores magnetic forces, which
should be of little importance for neutral atoms. If appreciable
magnetic fields are present, however, ions with net charges may
be significantly affected. We ignore this complication for the
moment and assume that the magnetic fields are small enough to
play a negligible role for the kinematics of the observed gas
particles in the disk.

For $\beta$\,Pic we estimate $\beta_{\mathrm{Na\,I}} = 250$ and
$\beta_{\mathrm{Fe\,I}} = 18$, while $C$ depends on the detailed
density and temperature structure. We calculated terminal velocities as a function
of radius for the two models of
Sects.~\ref{lightsect}~\&~\ref{heavysect}. Details of the
solution to Eq.\,\ref{motioneq}, with estimates of $C$ and $\beta$
for several ions, are found in \citet{lis03}.

\subsection{Ionisation structure}
\label{cloudy}
To estimate the gas density of all Na atoms in the disk, and also
get an idea of the total gas density, we need to address the
ionisation structure of the disk. We employed the one-dimensional
photoionisation/PDR code {\it Cloudy} \citep{fer98} for the task. 
Cloudy consistently maintains the ionization and thermal balance,
and solves the radiative transfer by making use of the Sobolev
approximation. For a detailed description of the code we used,
see \citet{lis99}.

Several assumptions have to be made in order to
compute the ionisation structure, perhaps the most important
being that of the chemical abundances of the atomic gas. 
Since we have no detailed information, we confine ourselves
to study two cases of chemical abundances: solar composition
and a strongly metal depleted model. At first we also considered
a case with typical interstellar medium abundances, but
dismissed it due to the very high degree of Ca depletion
compared to solar values, in contradiction with observations.
We have no {\it a priori} reason to expect any of these assumed
abundances to reflect reality, but we believe they still serve
as valuable reference cases.

Another important assumption is that of the stellar spectral
energy distribution (SED). We assume the atmosphere of
$\beta$\,Pic to be well represented by an ATLAS\,9 model \citep{kur92}
with $T_{\mathrm{eff}}=8000$\,K, $\log g = 4.5$ (in cm\,s$^{-2}$)
and $\log(Z/Z_{\odot})=0.0$, and the effective radius to be
$R = 1.75 R_{\odot}$, implying the luminosity $L = 11 L_{\odot}$.
Recent VLTI observations of $\beta$\,Pic estimate the stellar
radius to be $(1.735\pm0.128) R_{\odot}$ \citep{dif03}, and
comparing the ATLAS\,9 SED of $\beta$\,Pic with observed data
from the HST and FUSE archives, we find the agreement to be
generally excellent \citep[see][]{lis03}. Some deviations are found in
the far-ultraviolet, where $\beta$\,Pic seems to be a slightly
atypical A5V star with possible chromospheric activity
\citep{bou02}.

\subsubsection{Solar composition disk}
\label{lightsect}
Assuming solar abundances throughout the disk, we
can fit the radial Na\,I density of Sect.~\ref{denssect}
to within 1\,\% by setting the radial density profile
of hydrogen nucleii to
\begin{equation}
\label{lighteq}
n(\mathrm{H})=2.25\times10^3 \left[ \left( \frac{r}{r_0} \right)^{2.4} +
\left( \frac{r}{r_0} \right)^{5.3} \right]^{-\frac{1}{2}}\mathrm{cm}^{-3},
\end{equation}
where $r_0$ is the same as in Sect.~\ref{denssect}. The average
fraction of neutral Na in the model is $10^{-3}$, while hydrogen is
entirely neutral. The hydrogen column densities are $N(\mathrm{H\,I}) =
8\times10^{18}$\,cm$^{-2}$ and $N(\mathrm{H_2}) =
3\times10^{18}$\,cm$^{-2}$, consistent with the observational
upper limits of $N(\mathrm{H\,I}) \la$ a few $\times 10^{19}$\,cm$^{-2}$ \citep{fre95}
and $N(\mathrm{H_2}) \la 3 \times 10^{18}$\,cm$^{-2}$ 
\citep[$3\sigma$,][]{lec01}. The
gas temperature of the model disk is low, $T_{\mathrm{gas}}$(100\,AU) = 13\,K, while
the dust is significantly warmer, $T_{\mathrm{dust}}$(100\,AU) = 50\,K. 
With a radial density as in Eq.\,\ref{lighteq} and a vertical profile similar
to the Na\,I gas, the total gas disk mass out to 1\,000\,AU becomes
$\sim$0.1\,$M_{\oplus}$.

Using the equations of motion of Sect.~\ref{radpress} to
calculate the terminal velocity of elements subject to the
gas drag of the hydrogen density law of Eq.\,\ref{lighteq} and
temperature structure from Cloudy, we find that both Fe and Na,
both with $\beta \gg 1$, would reach
radial velocities on the order of hundreds to thousands km\,s$^{-1}$.
That is in clear contradiction with the observed radial velocities
reported in 
Table~\ref{radveltab}. The purported explanation of
Paper\,I, that Na is ionised most of the time and therefore is
subject to a small average radiation pressure, does not apply to Fe;
indeed, both Fe\,I and Fe\,II experience a radiation pressure with
$\beta\gg1$. We know of only one remaining plausible explanation:
there is a braking agent keeping the gas from reaching high
velocities.

\subsubsection{Metal depleted disk}
\label{heavysect}
To investigate the possibility of a more massive disk than implicated
by solar abundances of Na\,I, we impose a radial density profile for
hydrogen nucleii,
\begin{equation}
\label{heavyeq}
n(\mathrm{H})=10^6 \left[ \left( \frac{r}{r_0} \right)^{2.4} +
\left( \frac{r}{r_0} \right)^{5.4} \right]^{-\frac{1}{2}}\mathrm{cm}^{-3},
\end{equation}
and tune the metal depletion to reproduce the Na\,I density
profile. We found that an overall depletion factor of $\sim$10$^{-3}$
(more precisely $8.5 \times 10^{-4}$)
of the metals with respect to solar composition was required. Due
to higher dust-gas interaction, the temperatures of the gas and
the dust are similar over the disk, with
$T_{\mathrm{gas}}$(100\,AU) = 36\,K and $T_{\mathrm{dust}}$(100\,AU) = 51\,K.
The column densities of hydrogen are $N(\mathrm{H\,I}) =
6\times10^{20}$\,cm$^{-2}$ and $N(\mathrm{H_2}) =
3\times10^{21}$\,cm$^{-2}$, significantly higher than the
observed upper limits reported by \citet{fre95} and
\citet{lec01}. With a radial density as in
Eq.\,\ref{heavyeq} and a vertical profile similar to the Na\,I
gas, the total gas disk mass out to 1\,000\,AU becomes
$\sim$40\,$M_{\oplus}$, similar to the mass recently inferred
from possible emission in the pure rotational transitions of H$_2$,
$J=2$$\rightarrow$$0$ (28\,$\mu$m) and $J=3$$\rightarrow$$1$
(17\,$\mu$m), detected from $\beta$\,Pic in ISO/SWS data
\citep{thi01}. Due to the low gas temperature of our model,
however, the predicted fluxes in the 17\,$\mu$m
and 28\,$\mu$m lines are only
$F_{17} = 6 \times 10^{-15}$\,erg\,s$^{-1}$\,cm$^{-2}$ and
$F_{28} = 2 \times 10^{-15}$\,erg\,s$^{-1}$\,cm$^{-2}$, respectively, 
for ortho/para in thermal equilibrium.
This is more than an order of magnitude lower than the fluxes
$F_{17} = 7.7 \times 10^{-14}$\,erg\,s$^{-1}$\,cm$^{-2}$ and
$F_{28} = 7.0 \times 10^{-14}$\,erg\,s$^{-1}$\,cm$^{-2}$
reported by \citet{thi01}, and can be traced to their higher
assumed gas temperature of 100\,K.

\begin{figure}
 \resizebox{\hsize}{!}{\includegraphics{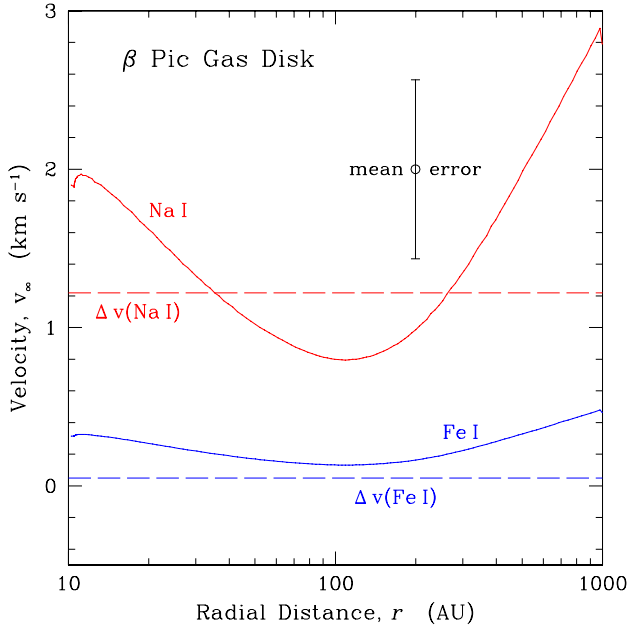}}
 \caption{The terminal velocities for Na\,I and Fe\,I in
a metal depleted disk with the radial density of
Eq.\,\ref{heavyeq}, as a function of radius. The horizontal
dashed lines show the measured radial velocities,
as determined from absorption lines (Table~\ref{radveltab}),
with the estimated error shown by the vertical bar. The error
is dominated by the wavelength calibration uncertainty of
0.5\,km\,s$^{-1}$.}
 \label{termvelfig}
\end{figure}

Calculating the terminal velocities of the radiation pressure
sensitive elements Fe and Na, we produce Fig.~\ref{termvelfig},
where the terminal velocity is plotted against the radial distance.
We find that the derived terminal velocities are consistent with
observed radial velocities reported in Table~\ref{radveltab}.

\subsection{Summary}
We have found that in our thin disk, constructed by 
fitting a solar composition gas to the Na\,I observations,
the predicted H\,I and H$_2$ column densities are consistent
with the observed limits. With these low densities, however, the
radiation pressure is predicted to accelerate, e.g., Na\,I to
radial velocities inconsistent with observations. 

On the other hand, in our thick disk, constructed by fixing
the H density to be consistent as the braking agent for the
metals, the predicted column densities of H\,I and H$_2$ are
both above the observed upper limits, although the predicted
H$_2$ {\it emission} is still far below the claimed detection.

The situation is puzzling, but a possible solution may be found 
if a main braking agent {\em different} from hydrogen is in action.
If, e.g., the observed gas is replenished from dust collisions rather than
being primordial, as seems suggested by the close spatial correlation
between gas and dust, then a more natural candidate might be 
oxygen or oxygen bearing species (P. Artymowicz, private communication).
More detailed models are in preparation, but hopefully the upcoming Space
InfraRed Telescope Facility will solve the issue unambiguously by better
observational constraints.

\section{Conclusions}
Our main observational results are:
   \begin{enumerate}
   \item We have observed 88 spatially resolved emission lines coming from 
     Fe\,I, Na\,I, Ca\,II, Ni\,I, Ni\,II, Ti\,I, Ti\,II, Cr\,I and
     Cr\,II in the $\beta$\,Pictoris gas disk.
   \item We trace the gas emission to the limits of our observations, from 0\farcs7
     (13\,AU) out to 17\arcsec\ 
     (323\,AU) radially to the NE in Na\,I, and 4\arcsec\ (77\,AU) above the disk
     plane at radius 6\arcsec\ (116\,AU) in Ca\,II.
   \item The scale height of the gas at 6\arcsec\ from the star is twice as high as
     the equivalent dust scale height.
   \item There is a brightness NE/SW asymmetry in the gas emission reminiscent
     of the dust asymmetry, although much stronger.
   \item The inner gas disk is tilted by $\sim$5\degr, similarly to the dust disk. 
   \item The heliocentric radial velocity of $\beta$\,Pic is 20.0$\pm$0.5\,km\,s$^{-1}$.
   \item The radial velocities of ions observed in absorption are close to or
     at the systematic velocity of $\beta$\,Pic (to a few km\,s$^{-1}$).
   \item The estimated radial density from Na\,I in emission predicts a column
     density similar to the one observed in absorption, meaning that most, perhaps
     all, Na\,I is distributed in the observed disk. We have no reason to doubt 
     that originators of other 'stable' gas absorption components also
     belong to this extended disk.
   \item Our disk models show that assuming hydrogen to be the braking agent
     for metals pushed out by radiation pressure in the $\beta$\,Pic disk
     leads to contradictions with observations. 
   \end{enumerate}
A more detailed study of the chemical composition of the
$\beta$\,Pic disk will be presented in a forthcoming paper.

\begin{acknowledgements}
  We would like to thank the staff at the VLT and UVES for their outstanding
contributions in performing these demanding service observations, in particular
Fernando Comer\'{o}n for kindly rescheduling an erroneous observation block, and
Andrea Modigliani for addressing errors in the UVES pipeline. We acknowledge 
the interesting discussions we have had with Pawel Artymowicz, Doug Lin, and Philippe
Th\'{e}bault. We thank the referee Alain Lecavelier des Etangs for a rapid and
detailed report.
This research has made use of NASA's Astrophysics Data System Bibliographic Services,
atomic line lists compiled by the National Institute for Standards and Technology,
and the SIMBAD database.
\end{acknowledgements}

\bibliographystyle{aa}
\end{document}